\documentclass[12pt]{article}
\usepackage[utf8]{inputenc}
\usepackage{hyperref}
\usepackage{amsmath}
\usepackage{graphicx}
\usepackage{enumitem}

\title{\vspace{-2cm}Agentic Systems: A Guide to Transforming Industries with Vertical AI Agents\vspace{1cm}}

\author{
    Fouad Bousetouane\textsuperscript{1,2} \\[1em] 
    \textsuperscript{1}\text{The University of Chicago, USA} \\[0.75em] 
    \textsuperscript{2}\text{2ndsight.ai} \\[1.5em] 
    {\small \href{mailto:bousetouane@uchicago.edu}{\texttt{bousetouane@uchicago.edu}}}
}

\date{}

\begin{document}

\maketitle

\begin{abstract}
The evolution of agentic systems represents a significant milestone in artificial intelligence and modern software systems, driven by the demand for vertical intelligence tailored to diverse industries. These systems enhance business outcomes through adaptability, learning, and interaction with dynamic environments. At the forefront of this revolution are Large Language Model (LLM) agents, which serve as the cognitive backbone of these intelligent systems. 

In response to the need for consistency and scalability, this work attempts to define a level of standardization for Vertical AI agent design patterns by identifying core building blocks and proposing a \textbf{Cognitive Skills } Module, which incorporates domain-specific, purpose-built inference capabilities. Building on these foundational concepts, this paper offers a comprehensive introduction to agentic systems, detailing their core components, operational patterns, and implementation strategies. It further explores practical use cases and examples across various industries, highlighting the transformative potential of LLM agents in driving industry-specific applications.
\end{abstract}

\newpage
\tableofcontents
\newpage

\section{Introduction}
The rapid evolution of technology has transformed business operations, with SaaS platforms \cite{AIforITOps2023} becoming essential for scalability and efficiency across industries. However, as industries face increasingly dynamic and complex environments, traditional SaaS solutions often fall short in meeting domain-specific and evolving needs.

To bridge this gap, \textbf{agentic systems} have emerged as a new generation of solutions. Powered by LLMs and advanced AI capabilities, they deliver intelligent, context-driven, and domain-specific solutions, addressing the limitations of both traditional SaaS platforms and context-aware systems.

\subsection{The Shortcomings of Traditional SaaS Platforms}
Traditional SaaS platforms serve as the backbone of business operations, offering reliable tools for managing workflows and maintaining operational consistency. Their architecture emphasizes horizontal scalability and general applicability, enabling businesses to standardize processes and optimize routine tasks across industries. This broad applicability makes SaaS ideal for managing repetitive tasks and scaling operations across diverse sectors. However, this generalized design often comes at the expense of domain-specific intelligence and flexibility, which are critical for addressing the unique challenges of dynamic and complex environments.

These limitations are evident in various industries:
\begin{itemize}
    \item \textbf{E-commerce:} Platforms efficiently handle online transactions, product catalog management, and order tracking. Yet, they often require extensive customization to analyze customer purchasing behaviors, predict seasonal demand trends, or dynamically adjust inventory levels based on real-time sales data.
    \item \textbf{Multichannel Marketing:} Tools streamline campaign management across various channels, offering templates and automation for email, social media, and advertisements. However, their reliance on predefined workflows limits their ability to adapt quickly to shifting customer preferences, emerging trends, or competitor strategies.

\newpage
    \item \textbf{Inventory Management:} Systems track stock levels and trigger reorders based on predefined thresholds. Despite this, they typically lack the ability to anticipate supply chain disruptions, respond to sudden demand spikes, or optimize procurement strategies using external market insights.
\end{itemize}

These examples underscore the reliance of traditional SaaS platforms on rule-based automation and structured data inputs. While effective for predictable and routine processes, they fall short in addressing domain-specific tasks that require contextual intelligence and adaptability.

\subsection{The Transition to Context-Aware Systems}
The limitations of traditional SaaS platforms have driven the adoption of \textbf{context-aware systems}, which aim to address these gaps by integrating real-time data and adaptability into workflows. By dynamically adjusting to evolving scenarios, these systems enable businesses to operate more effectively in increasingly complex environments. Context-aware systems are designed to:
\begin{itemize}
    \item \textbf{Understand dynamic environments:} Incorporate real-time data to adjust workflows and outputs.
    \item \textbf{Bridge data to decisions:} Translate raw data into actionable insights without extensive manual intervention.
    \item \textbf{Adapt to evolving scenarios:} Adjust to unforeseen conditions or emerging trends.
\end{itemize}

While these systems represent a significant step forward, they still face challenges. For example:
\begin{itemize}
    \item \textbf{Supply Chain Management:} Traditional tools track inventory but fail to predict disruptions caused by external factors such as weather events or geopolitical risks.
    \item \textbf{Healthcare:} Scheduling systems can manage appointments but lack the capability to prioritize critical patients based on real-time health data.
\end{itemize}

These examples highlight the limitations of context-aware systems in achieving comprehensive decision-making. While they offer adaptability, they are constrained by their dependence on predefined rules and lack the ability to process unstructured data or make advanced contextual decisions.

\section{The Rise of Vertical AI Agent Solutions}

As industries face increasingly complex and domain-specific challenges, the limitations of traditional and context-aware systems have become evident. Vertical AI agents have emerged as a transformative solution, embedding industry-specific expertise and fine-tuned intelligence into adaptable, real-time systems. By combining the flexibility of context-aware systems with domain knowledge, they empower organizations to address unique challenges with precision and efficiency.

These agents bridge the gap between general-purpose systems and the specific demands of modern industries, enabling real-time adaptability and specialized problem-solving. This evolution marks a pivotal shift in intelligent system design, allowing businesses to optimize workflows, enhance decision-making, and tackle increasingly dynamic operational requirements with unprecedented effectiveness.

\subsection{Operational Advantages of Vertical AI Agents}

\subsubsection{1. Targeted Domain Expertise}
Vertical AI agents are tailored for specific industries, utilizing domain-specific reasoning engines (LLMs) fine-tuned for specialized knowledge and workflows to address complex challenges effectively. This ensures they can:
\begin{itemize}
    \item Perform intricate tasks, such as legal contract analysis, medical imaging interpretation, or financial risk assessment, with exceptional precision.
    \item Generate insights and recommendations tailored to the unique demands of the domain, reducing errors and manual effort.
    \item Ensure operational accuracy and alignment with industry standards by incorporating domain-specific protocols and guidelines directly into their decision-making processes, minimizing risks and errors in critical tasks.
\end{itemize}
These specialized capabilities make vertical AI agents indispensable in fields where accuracy, reliability, and regulatory adherence are critical.

\subsubsection{2. Dynamic Adaptability in Real-Time Operations}
Unlike traditional systems, vertical AI agents excel in dynamic environments, continuously adapting to changing conditions and operational demands. They achieve this through:
\begin{itemize}
    \item \textbf{Real-Time Data Processing:} Leveraging live inputs like inventory fluctuations, customer preferences, or environmental factors to adjust strategies and outputs instantly.
    \item \textbf{Proactive Decision-Making:} Anticipating disruptions and reconfiguring workflows, such as rerouting supply chains during delays or reallocating resources in emergencies.
    \item \textbf{Scalable Responsiveness:} Managing both minor adjustments and large-scale shifts with agility, ensuring minimal downtime and maximum efficiency.
\end{itemize}
This adaptability empowers organizations to respond effectively to evolving challenges, making vertical AI agents central to resilient and responsive operations.

\subsubsection{3. End-to-End Workflow Automation}
By automating complex processes, vertical AI agents transform raw data into actionable outcomes, streamlining workflows traditionally reliant on human intervention. This results in:
\begin{itemize}
    \item \textbf{Faster Turnaround Times:} Analyzing, deciding, and executing tasks within seconds, significantly reducing delays in processes like customer onboarding or compliance reviews.
    \item \textbf{Cost Optimization:} Automating repetitive tasks allows human resources to focus on strategic, high-value activities, increasing productivity and reducing operational expenses.
    \item \textbf{Interoperability Across Systems:} Seamlessly integrating with enterprise tools and bridging gaps between structured (e.g., ERP systems) and unstructured (e.g., emails, documents) data environments.
\end{itemize}

Vertical Ai agent solutions are rapidly gaining momentum, with major players such as Google, AWS, OpenAI, and Microsoft spearheading efforts to develop platforms that simplify and scale the creation of vertical AI solutions. While these advancements signal a transformative shift, we are still in the early stages of this journey, with operational patterns only beginning to take shape. These emerging platforms aim to provide standardized frameworks for fine-tuning, deployment, and integration, enabling a more structured approach to building intelligent, adaptive agents. In \textbf{Section 3}, we explore \textbf{LLM agents}, the bedrock of vertical AI agents, leveraging large language models for domain-specific intelligence and adaptability. \textbf{Section 4} introduces \textbf{agentic systems}, their categories, operational patterns, and transformative industry applications.

\section{What Are LLM Agents?}
\subsection{Definition}
\begin{figure}[h!]
    \centering
    \includegraphics[width=\textwidth]{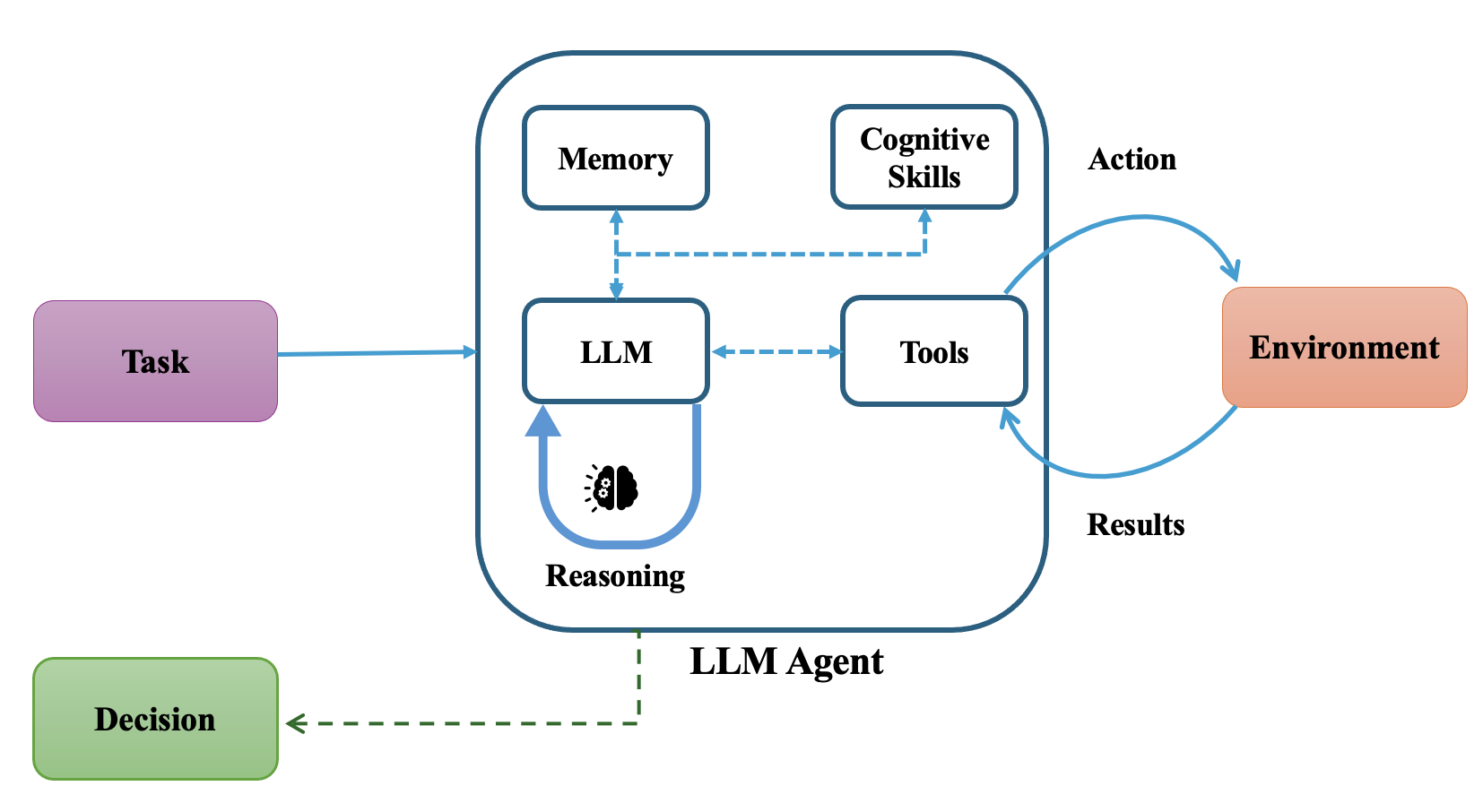}
    \caption{Architecture and Core Components of an LLM Agent}
    \label{fig:architecture}
\end{figure}

LLM agents are autonomous, intelligent systems powered by Large Language Models (LLMs) that integrate modular components—reasoning, memory, cognitive skills, and tools—to solve complex tasks in dynamic and evolving environments. These agents are designed to operate independently, adapt to changes, and execute sophisticated tasks by combining domain-specific expertise with contextual understanding. 
Each module within the agent's architecture serves a distinct purpose: reasoning enables logical decision-making, memory supports retention and recall of critical information, and tools facilitate interaction with external systems and environments. Figure \ref{fig:architecture} illustrates the modular architecture and components of an LLM agent, highlighting its ability to perform dynamic, real-time processes with adaptability, intelligence, and precision.

We introduce a new module to the core building blocks of the LLM agent—\textbf{cognitive skills}—which fills the gap between pre-trained or fine-tuned LLM reasoning, external tools for interacting with the environment, and new inference models. This module ensures that LLM agents are equipped with purpose-built models tailored to specific tasks, enhancing their ability to operate effectively across various domains and challenges.

\subsection{LLM Agents vs. LLM Workflows}
It is important to distinguish LLM agents from LLM workflows, as they differ both conceptually and operationally. LLM workflows are predefined, static processes designed to perform specific, linear tasks. They operate based on a structured pipeline where each step is explicitly defined and executed in sequence, with little to no flexibility or adaptability.

\begin{figure}[h!]
    \centering
    \includegraphics[width=\textwidth]{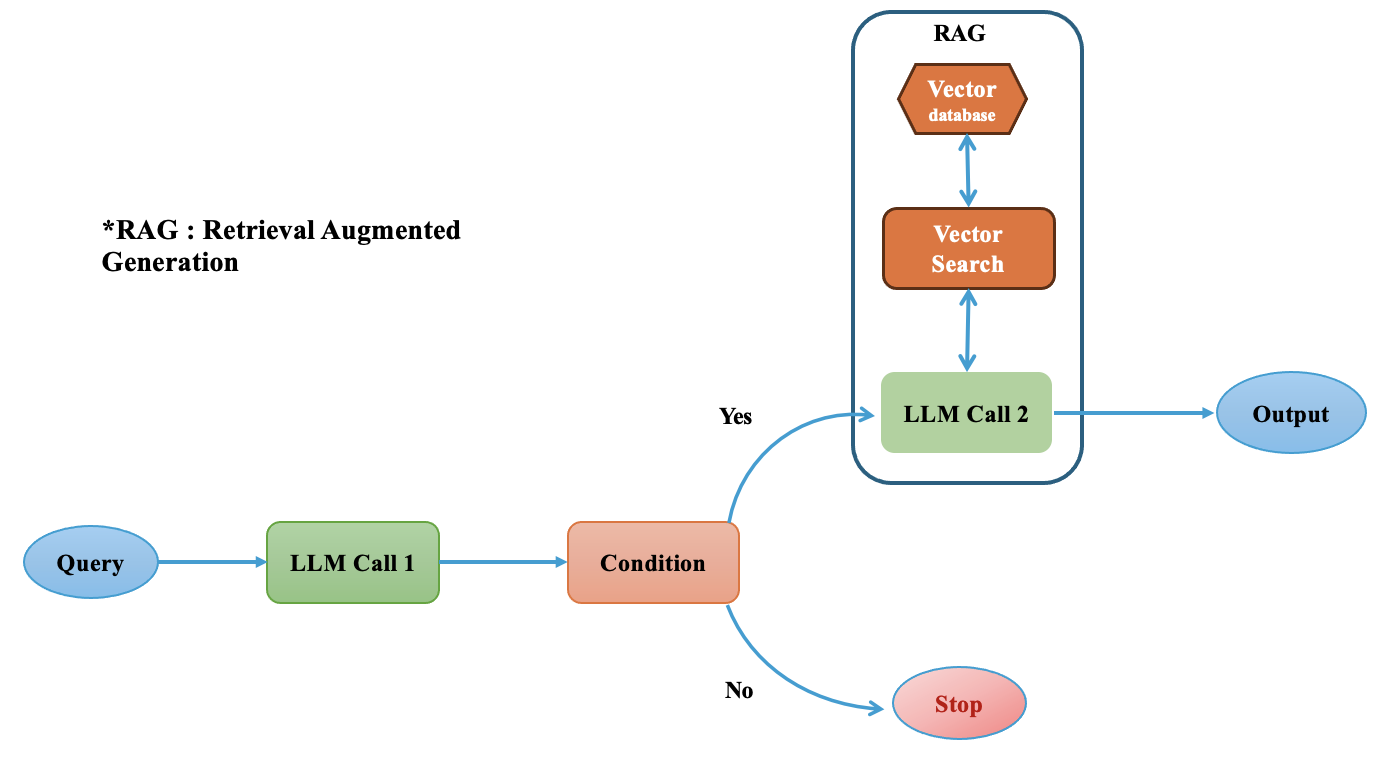}
    \caption{Example of LLM Workflow: Chain Prompting with RAG for Knowledge Retrieval}
    \label{fig:workflow}
\end{figure}

For instance, as illustrated in Figure \ref{fig:workflow}, a typical workflow involves a chain of prompts using multiple LLMs, combined with a Retrieval-Augmented Generation (RAG) pattern for accessing domain-specific knowledge. In this setup, one LLM might process the query to determine intent or refine context, while another LLM, equipped with retrieved knowledge, generates the final response. The workflow’s reliance on fixed steps ensures consistency but limits flexibility. For more details on RAG implementation and advanced prompting guidelines, refer to \cite{giray2023prompt} and \cite{gao2023retrieval}.

LLM agents stand apart due to their ability to reason, adapt, and refine their actions in response to changing environments and complex goals, making them well-suited for advanced, dynamic applications. This distinction underscores the versatility and intelligence that define LLM agents as compared to traditional LLM workflows.

\newpage
\subsection{Core Modules of LLM Agents}

\subsubsection{Memory: The Core of Continuity and Context}
The memory module underpins the agent’s ability to maintain context across interactions, ensuring personalized and consistent responses. It stores historical interactions, user preferences, and domain-specific knowledge, serving as the agent’s long-term storage system. By leveraging memory, the agent achieves:
\begin{itemize}
    \item \textbf{Contextual Awareness:} Drawing on prior interactions to maintain continuity.
    \item \textbf{Personalization:} Adapting responses based on user-specific information.
    \item \textbf{Domain Expertise:} Utilizing stored knowledge to deliver precise and informed outputs.
\end{itemize}
This module ensures that the agent operates seamlessly, integrating past interactions with real-time data to provide contextually appropriate results.

\subsubsection{Reasoning Engine (LLM): The Brain of the Agent}
The Reasoning Engine module, powered by the LLM, is the decision-making core of an LLM agent. It orchestrates logical inference, planning, contextual understanding, and personalized interaction, transforming raw data into actionable insights. By integrating inputs from Memory, Cognitive Skills, and Tools, the Reasoning Engine ensures the agent operates effectively in dynamic and complex environments. As illustrated in Figure \ref{fig:architecture}, this module lies at the heart of agentic intelligence, driving coherence and adaptability in every interaction.

\paragraph{Core Capabilities of the Reasoning Engine}
\begin{enumerate}
    \item \textbf{Logical Inference and Problem-Solving:} The Reasoning Engine evaluates inputs to derive meaningful conclusions. By analyzing ambiguous or complex scenarios, it applies advanced logical reasoning to ensure the agent’s responses are precise and data-driven.
    \item \textbf{Contextual Understanding and Response Generation:} Leveraging historical data from Memory and real-time inputs from Tools, the Reasoning Engine tailors outputs to the context, ensuring coherent, adaptive, and aligned interactions. This contextual understanding enables the agent to handle diverse and evolving scenarios with accuracy.
    \item \textbf{Task Sequencing, Goal-Oriented Planning, and Chain of Thought Reasoning:} The Reasoning Engine strategically organizes and sequences tasks, ensuring goal-oriented behavior. A key enhancement is its Chain of Thought Reasoning, which allows the agent to break down complex queries into smaller, sequential steps. This process ensures clarity, logical flow, and accurate resolution, especially for multifaceted tasks.
    \item \textbf{Adaptive Personas for Tailored Interactions:} The Reasoning Engine integrates personas to adapt the agent’s tone, style, and reasoning approach based on its audience. Personas enhance user trust and engagement by aligning interactions with expectations:
    \begin{itemize}
        \item \textbf{Empathetic Persona:} Suitable for healthcare or customer support, offering compassionate and understanding interactions.
        \item \textbf{Professional Persona:} For business or legal applications, ensuring formal and precise responses.
        \item \textbf{Casual Persona:} For consumer-facing roles, promoting friendly and approachable communication.
    \end{itemize}
\end{enumerate}

\subsubsection{Cognitive Skills: Task-Specific Inferences}

The Cognitive Skills module acts as a \textbf{model hub}, equipping the agent with purpose-built models specifically designed to accomplish tasks that general-purpose LLMs, even when fine-tuned LLMs, struggle to perform effectively. Fine-tuned models often lack the precision and specialization required for complex, domain-specific tasks. The Cognitive Skills module bridges this gap by delivering domain-specific cognitive capabilities uniquely tailored for specialized applications, enhancing the agent’s functionality and adaptability. By leveraging these specialized skills, the agent can tackle tasks requiring high precision, domain expertise, or advanced processing capabilities.
\\

\textbf{Examples of Cognitive Skills in Action:}

\begin{itemize}
    \item \textbf{Risk Assessment Models:}  
    Built for screening and evaluating:
    \begin{itemize}
        \item Intellectual property content for potential infringement or conflicts.
        \item Personal information and privacy-sensitive data to ensure compliance with regulations such as GDPR.
        \item Risk-prone operations, such as credit underwriting in financial services.
    \end{itemize}

    \item \textbf{Vulnerability Detection Models:}  
    Developed to protect against adversarial attacks and vulnerabilities by:
    \begin{itemize}
        \item Identifying and mitigating risks like jailbreaking attempts, toxic content generation, or data poisoning attacks.
        \item Enhancing the agent's resilience in adversarial environments, ensuring reliable performance.
    \end{itemize}
\newpage
    \item \textbf{Compliance Monitoring Models:}  
    Critical for ensuring adherence to:
    \begin{itemize}
        \item Organizational policies by detecting outputs that deviate from ethical or operational guidelines.
        \item Legal frameworks, such as validating contracts or outputs against industry-specific regulations.
    \end{itemize}

    \item \textbf{Optical Character Recognition (OCR):}  
    Enables the agent to process and extract information from:
    \begin{itemize}
        \item Scanned documents, invoices, or receipts.
        \item Handwritten forms or images containing text.
        \item Complex documents requiring structured data extraction.
    \end{itemize}

    \item \textbf{Image Classification and Object Detection:}  
    Provides visual processing capabilities, such as:
    \begin{itemize}
        \item Identifying defective parts in manufacturing processes.
        \item Classifying medical images for diagnostics (e.g., detecting tumors in X-rays).
        \item Analyzing satellite imagery for environmental monitoring.
    \end{itemize}

    \item \textbf{Audio and Speech Processing Models:}  
    Adds specialized capabilities to handle:
    \begin{itemize}
        \item Transcription and sentiment analysis of call center recordings.
        \item Real-time language translation in multilingual communication.
    \end{itemize}

    \item \textbf{Responsible AI - Guardrail Classifiers:}  
    Essential for ensuring ethical and safe agent operations, including:
    \begin{itemize}
        \item \textbf{Toxicity Detection:} Screening outputs for offensive or harmful language.
        \item \textbf{Bias Mitigation:} Identifying and reducing biases in generated responses.
        \item \textbf{Ethical Review:} Validating outputs for alignment with societal and organizational ethical standards.
        \item \textbf{Misinformation Detection:} Flagging and correcting potentially false or misleading information.
    \end{itemize}
\end{itemize}

These cognitive skills enable the agent to function as a \textbf{versatile, purpose-driven system} that adapts to its operational environment by leveraging specialized capabilities. By acting as a \textbf{bridge between the LLM’s general reasoning abilities and domain-specific inference tasks}, this module ensures that the agent is not only adaptable but also precise, reliable, and aligned with industry-specific needs.

\subsubsection{Tools: Bridging Knowledge and Interaction}
The Tools module equips the agent with a range of tools that enhance its ability to be contextually and environmentally aware. These tools enable the agent to access, retrieve, and process information from various sources, ensuring its actions are informed, adaptive, and aligned with operational goals. Examples of tools the agent can leverage include:
\begin{itemize}
    \item \textbf{Knowledge Retrieval Systems:} Retrieval-Augmented Generation (RAG) systems to access structured (e.g., databases) and unstructured (e.g., document repositories) knowledge, enabling the agent to incorporate relevant domain-specific information into its operations.
    \item \textbf{Dynamic API Integration:} Tools that allow the agent to interact with live data streams, proprietary platforms, and external systems, facilitating real-time decision-making and adaptive responses.
    \item \textbf{Legacy System Interfaces:} Tools for bridging traditional structured data systems, such as relational databases, to incorporate historical data and insights into the agent's current tasks.
    \item \textbf{Contextual Awareness Tools:} Systems that provide the agent with situational and environmental context, enabling it to tailor its actions and outputs based on specific operational scenarios.
\end{itemize}

\newpage

\section{Agentic Systems: Definition, Categories, and Applications}

\subsection{Definition of Agentic Systems}
Agentic systems are advanced frameworks that integrate one or more LLM agents to automate complex tasks and streamline processes across various domains. These systems are designed to function autonomously, enabling agents to collaborate through direct communication or an orchestration module that coordinates their interactions. By leveraging modular designs, agentic systems provide flexibility, adaptability, and scalability to address dynamic and evolving operational needs.

\subsection{Architectural Flexibility and Design Patterns}
At the time of writing this article, there are no universally accepted design patterns for agentic systems. Their architectures and implementations vary significantly, often being tailored to specific domains and use cases. This flexibility allows organizations to design agentic systems that best align with their unique requirements, though it also presents challenges in establishing standardization and interoperability.

Despite the absence of standardized design patterns, substantial progress has been made by technology leaders, LLM solution providers, and academic researchers in advancing agentic frameworks. These efforts focus on developing generic frameworks while also building multi-agent systems tailored to domain-specific applications.

\subsubsection{Industry Efforts in Agentic Frameworks}
\begin{itemize}
    \item \textbf{Microsoft:} Introduced frameworks like AutoGen, which supports multi-agent systems for task automation and collaboration, and Semantic Kernel, which integrates AI into enterprise workflows with a focus on security and scalability \cite{microsoft_agentic}.
    \item \textbf{OpenAI:} Introduced the Assistants API, enabling the development of AI agents with advanced capabilities such as tool utilization, memory-based persistent conversations, and knowledge retrieval for handling complex tasks and dynamic interactions \cite{openai_devday_2023}.
    \item \textbf{Google:} Developed Vertex AI Agent Builder, which integrates Vertex AI Search for grounded responses and Vertex AI Conversation for natural dialogue, streamlining the development of agents for tasks like customer support and data analysis \cite{google_vertex}.
    \item \textbf{Amazon Web Services (AWS):} Provides a robust suite of tools tailored for deploying agentic systems across industries, allowing developers to address specific use cases \cite{aws_llm_agents}.
    \item \textbf{Anthropic:} Focuses on creating diverse agentic system patterns leveraging its core LLM, Claude, for various applications \cite{anthropic}.
    \item \textbf{LangChain:} Supports implementing agents for dynamic, multi-step tasks but faces challenges with speed limitations when managing complex interactions between multiple agents and tools \cite{langchain_agents}.
\end{itemize}

\subsubsection{Academic Research Efforts}
\begin{itemize}
    \item \textbf{Magentic-One:} Proposes a generalist multi-agent system architecture for solving complex problems, aiming for adaptability across domains \cite{MagenticOne2024}.
    \item \textbf{KG4Diagnosis:} Develops a hierarchical multi-agent framework enhanced with knowledge graphs to improve accuracy in medical diagnoses, particularly in healthcare \cite{KG4Diagnosis2024}.
    \item \textbf{MedAide:} Explores creating a collaborative medical assistant system using specialized LLMs to provide comprehensive patient support services \cite{MedAide2024}.
\end{itemize}
Together, these industry innovations and academic advancements are driving the rapid evolution of agentic systems, paving the way for more versatile and impactful AI solutions.

\newpage
\subsection{Categories of Agentic Systems}
Agentic systems can be categorized into three primary types based on their structure, scope, and interaction dynamics:
\begin{enumerate}
    \item \textbf{Task-Specific Agents}
    \item \textbf{Multi-Agent Systems}
    \item \textbf{Human-Augmented Agents}
\end{enumerate}
Each category reflects a unique approach to designing intelligent systems, tailored to address different operational needs and complexities.

\subsubsection{Task-Specific Agent}
\paragraph{Definition:}
A Task-Specific Agent is an autonomous system designed to handle a specific function or solve a narrowly defined problem within a particular domain. These agents act as specialized modules that contribute to larger systems by efficiently managing discrete tasks.

There are various patterns to implement Task-Specific Agents based on application needs. For example:
\begin{itemize}
    \item \textbf{ReAct Agent:} Combines reasoning and action to handle interactive workflows and decision-making tasks \cite{ReAct2022}.
    \item \textbf{Router Agent:} Maps queries or tasks to the appropriate sub-agents or data sources, often used in multi-domain retrieval systems like Retrieval-Augmented Generation (RAG) \cite{SRSA2024}.
\end{itemize}

In the next section, we will explore the architectural principles and use cases of the RAG Agent Router, a common implementation of the Router Agent pattern.

\newpage
\paragraph{RAG Agent Router :}

is a Task-Specific Agent designed to dynamically orchestrate knowledge retrieval in Retrieval-Augmented Generation systems. Its primary function is to analyze user queries and map them to the appropriate domain-specific knowledge sources, tools, or APIs, ensuring efficient and contextually accurate responses.

\begin{figure}[h!]
    \centering
    \includegraphics[width=\textwidth]{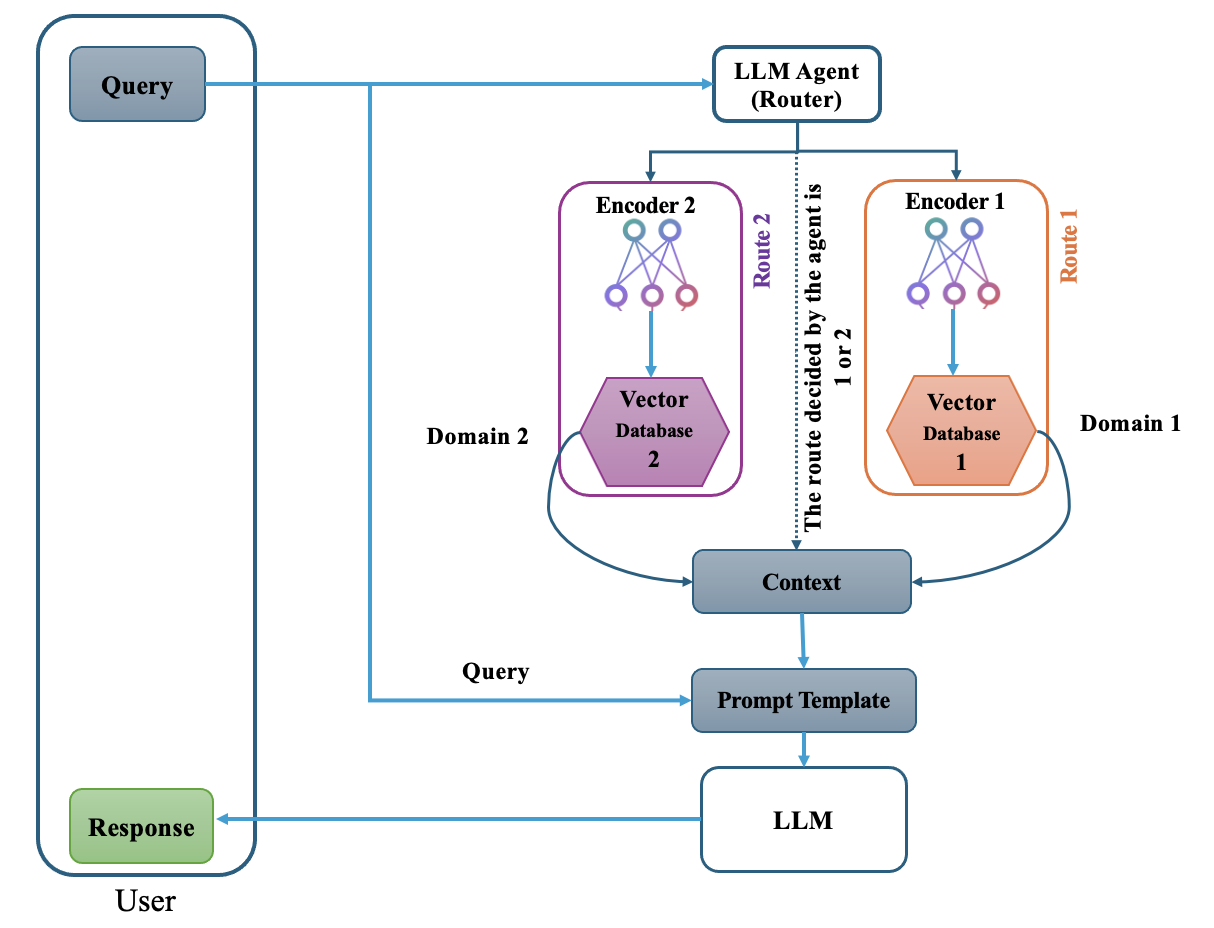}
    \caption{Architecture of the RAG Agent Router with Domain-Specific Vector Databases}
    \label{fig:rag_router}
\end{figure}

Figure \ref{fig:rag_router} illustrates the architecture of the RAG Agent Router. When a user submits a query, it is processed by the LLM Agent (Router), which determines the appropriate route based on the query's intent. The router maps the query to one of two distinct vector databases, each representing a specific knowledge domain (e.g., legal knowledge or financial data). These vector databases are powered by domain-specific encoders, fine-tuned to understand the semantics and key aspects of their respective domains. The relevant contextual information retrieved is combined with a prompt template and sent to the LLM, which generates a summarized and contextually accurate response. The response is then delivered back to the user, ensuring relevance and precision tailored to the query.

This pattern is particularly valuable in scenarios such as:
\begin{itemize}
    \item \textbf{Domain-Specific Knowledge Sources:} Where multiple vector databases are tailored to specific domains (e.g., legal, financial) and rely on fine-tuned encoders to understand the semantics and nuances of their respective fields.
    \item \textbf{Separation of Indexes:} When combining knowledge sources into a single index is impractical due to constraints like scalability, performance optimization, or security requirements.
    \item \textbf{Dynamic Query Handling:} When user queries vary in intent and context, requiring the router to apply specialized retrieval strategies to deliver accurate, domain-specific information.
    \item \textbf{Leveraging Different Tools or APIs:} When queries necessitate the use of specific external tools (e.g., calculators, data analysis APIs, or CRM integrations) to supplement retrieval with actionable insights or automated workflows.
\end{itemize}

\newpage
\paragraph{Practical Use Cases of the Router Agent}
\begin{enumerate}
    \item \textbf{Customer Support Systems}
 \begin{itemize}
    \item \textbf{Scenario:} A customer submits multiple queries spanning different domains, such as tracking shipments, processing returns, and requesting refunds.
    \item \textbf{Solution:} The Router Agent parses the intent of each query, maps them to the respective domain-specific knowledge or APIs (e.g., tracking database, returns system, payments database, customer relationship management (CRM) tools, inventory management systems, or live chat platforms), retrieves the necessary data, and aggregates the information into a cohesive response for the customer.
\end{itemize}
    \item \textbf{Enterprise Knowledge Management}
    \begin{itemize}
        \item \textbf{Scenario:} Employees need access to documents stored across multiple departmental databases, such as HR policies, legal precedents, and financial reports.
        \item \textbf{Solution:} The Router Agent identifies the domain of the employee’s query and routes it to the appropriate database. It retrieves the relevant document or data and presents it efficiently, ensuring fast and accurate access to information.
    \end{itemize}
    \item \textbf{Healthcare Decision Support}
    \begin{itemize}
        \item \textbf{Scenario:} A physician queries multiple systems to access patient history, lab results, and clinical guidelines for a specific medical condition.
        \item \textbf{Solution:} The Router Agent processes the query and maps it to the respective systems (e.g., electronic health records, diagnostic databases, and treatment guidelines). It retrieves the relevant information and integrates it into a single, comprehensive report for the physician.
    \end{itemize}
    \item \textbf{RAG Agent Router in Multi-Domain Retrieval}
    \begin{itemize}
        \item \textbf{Scenario:} A system with separate vector databases for legal, financial, and technical knowledge needs to process a user query such as: "Summarize recent IP law precedents in technology."
        \item \textbf{Solution:}
        \begin{enumerate}
            \item The Router Agent identifies the domain as "legal."
            \item It maps the query to the "Legal Precedents Database."
            \item It retrieves summaries of relevant cases and passes them to the LLM, which generates a concise and domain-accurate response tailored to the query.
        \end{enumerate}
    \end{itemize}
\end{enumerate}

\newpage
\subsubsection{Multi-Agent Systems}
\paragraph{Definition:}
A Multi-Agent System is a collection of autonomous agents designed to collaborate and solve interconnected problems or achieve shared goals. These systems act as distributed modules that work together by communicating and coordinating tasks, offering scalability and adaptability in complex workflows. Depending on the application, agents within the system may share a common memory or operate with separate, isolated memories to optimize task execution.

There are various patterns to implement Multi-Agent Systems based on application needs. For example:
\begin{itemize}
    \item \textbf{Orchestrated Multi-Agent System:} Involves a lead agent that delegates subtasks to specialized agents and integrates their outputs, commonly used in dynamic, multi-step workflows.
    \item \textbf{RAG Orchestrated Multi-Agent System:} Extends the orchestrated system by incorporating agents specialized in retrieval tasks, with each agent accessing a specific knowledge domain or tool. The lead agent dynamically routes queries to the relevant agents and integrates the retrieved information to ensure accurate and context-aware responses.
    \item \textbf{Collaborative Problem Solvers:} Agents communicate directly with one another to achieve shared objectives without central control. This pattern is suitable for decentralized or distributed tasks where agents share information to collectively solve problems.
\end{itemize}

In the next section, we will explore the architectural principles and use cases of the RAG Orchestrated Multi-Agent System, a common implementation of this pattern.

\newpage

\paragraph{RAG Orchestrated Multi-Agent System:}
is an advanced implementation of a Multi-Agent System where a lead agent coordinates the activities of multiple specialized agents, each focused on retrieval tasks from specific knowledge domains or tools. The lead agent acts as the central orchestrator, dynamically routing queries to the relevant retrieval agents, collecting their outputs, and integrating the information into a unified, context-aware response. This design ensures efficient handling of complex queries that require information from diverse, domain-specific sources. 

\begin{figure}[h!]
    \centering
    \includegraphics[width=\textwidth]{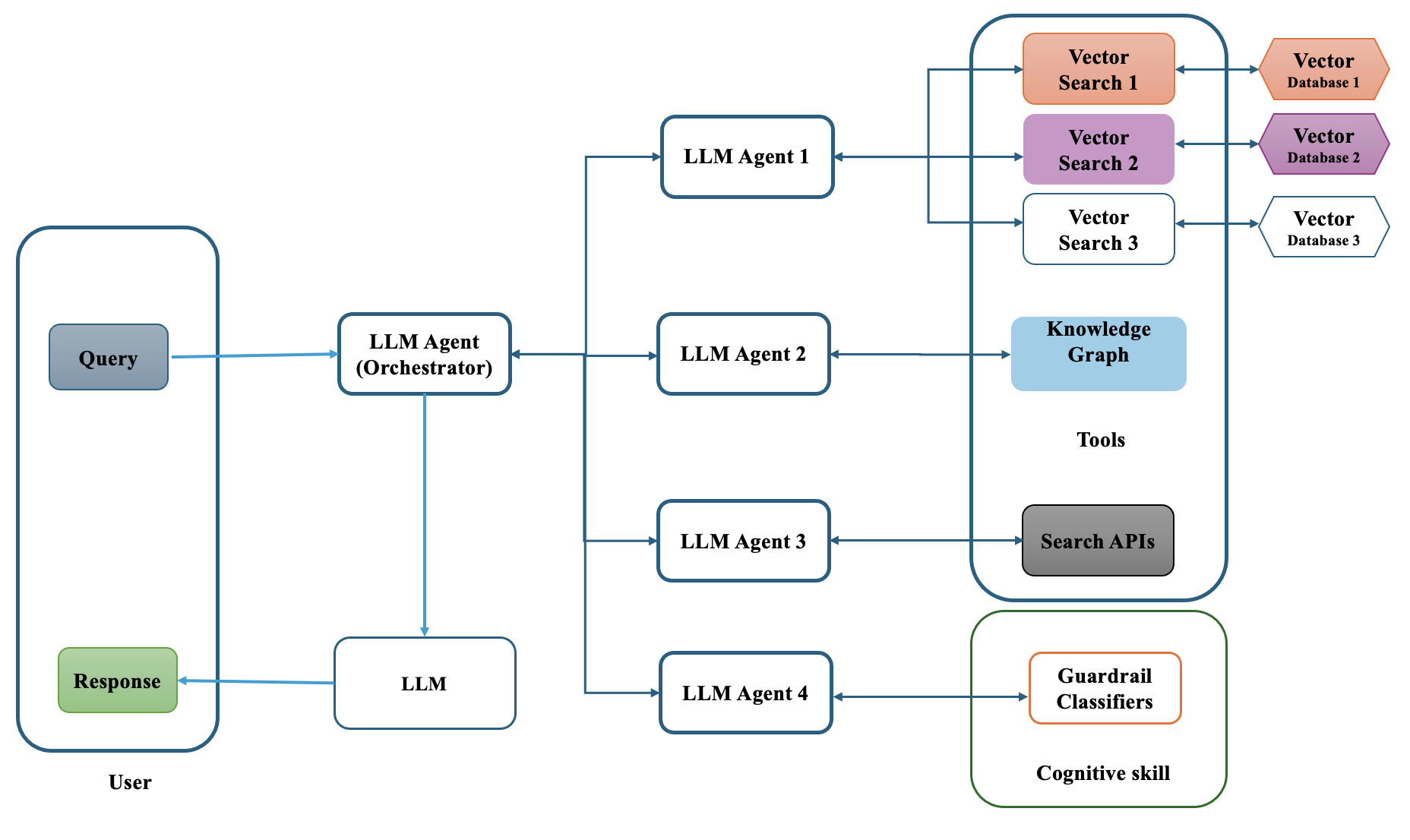}
    \caption{Architecture of the RAG Orchestrated Multi-Agent System for Multi-Domain Knowledge Retrieval}
    \label{fig:multi_agent_system}
\end{figure}

Figure \ref{fig:multi_agent_system} illustrates an example of a responsible RAG Orchestrated Multi-Agent System. When a user submits a query, it is first received by the LLM Agent (Orchestrator), which parses the query and determines how to decompose it into subtasks based on the query's intent. Each subtask is dynamically assigned to one of the specialized LLM Agents, which are responsible for interacting with distinct tools or cognitive skills.

This example highlights a specific architecture pattern for such a system, but additional agents can be integrated based on the application's requirements, providing flexibility for domain-specific or task-specific enhancements.

The specialized LLM Agents are connected to tools and cognitive skills, categorized as follows:

\begin{itemize}
    \item \textbf{Tools:}
    \begin{itemize}
        \item \textbf{Domain-Specific Sources:}
        \begin{itemize}
            \item \textbf{LLM Agent 1:} Connected to \textbf{Vector Search Engines}, which access specific vector databases (e.g., DB1, DB2, DB3). These databases represent unique knowledge domains, such as legal, financial, or technical data.
            \item \textbf{LLM Agent 2:} Connected to \textbf{Knowledge Graphs}, which provide structured and interconnected data for handling complex, interlinked queries.
        \end{itemize}
        \item \textbf{Broad Contextual Sources:}
        \begin{itemize}
            \item \textbf{LLM Agent 3:} Connected to \textbf{Search APIs}, leveraging external search engines or APIs to retrieve supplementary information and broader contextual data.
        \end{itemize}
    \end{itemize}
\newpage
    \item \textbf{Cognitive Skills:}
    \begin{itemize}
        \item \textbf{LLM Agent 4:} Utilizes \textbf{Guardrail Classifiers} to assess the risk levels of decisions made by the Orchestrator and other agents. These classifiers are pre-built to identify vulnerabilities, ethical concerns, and potential risks, ensuring that all outputs adhere to safety and responsibility guidelines.
    \end{itemize}
\end{itemize}

Each LLM Agent retrieves the necessary information from its assigned tool or skill, ensuring relevance and domain accuracy. The Orchestrator integrates the outputs from all agents, incorporating the risk assessment and validation performed by \textbf{LLM Agent 4}. The compiled context, along with the Orchestrator’s final decision, is then sent to the LLM, which processes this input to generate the final response. This response is delivered back to the user, ensuring it is cohesive, contextually accurate, and ethically sound.

\newpage

This pattern is particularly valuable in scenarios such as:
\begin{itemize}
    \item \textbf{Cross-Domain Information Retrieval:} When a query requires inputs from multiple specialized domains, such as combining legal precedents with financial data.
    \item \textbf{Dynamic Workflows:} Where queries need to be decomposed into subtasks that require different agents to retrieve or process information.
    \item \textbf{Scalable Knowledge Systems:} In systems with distributed or isolated knowledge bases, enabling retrieval without merging data into a single index.
    \item \textbf{Time-Sensitive Decision Support:} For example, providing executives with real-time insights by aggregating data from performance metrics, market analysis, and risk assessments.
\end{itemize}

\paragraph{Practical Use Cases of the RAG Orchestrated Multi-Agent System}
\begin{enumerate}
    \item \textbf{Enterprise Reporting}
    \begin{itemize}
        \item \textbf{Scenario:} A business executive requests insights on financial performance, customer feedback, and market trends.
        \item \textbf{Solution:}
        \begin{enumerate}
            \item The Orchestrator Agent splits the query into subtasks: financial analysis, customer sentiment, and market research.
            \item Each subtask is routed to specialized agents querying financial databases, sentiment analysis tools, and market research APIs.
            \item The outputs are integrated into a comprehensive report for the executive.
        \end{enumerate}
    \end{itemize}

\newpage
    \item \textbf{Healthcare Assistance}
    \begin{itemize}
        \item \textbf{Scenario:} A physician queries diagnostic criteria, patient history, and treatment options for a medical condition.
        \item \textbf{Solution:}
        \begin{enumerate}
            \item The Orchestrator Agent breaks the query into subtasks: diagnostics, patient history, and treatment plans.
            \item Specialized agents access diagnostic databases, EHRs, and clinical guidelines.
            \item Results are compiled into a single, detailed treatment recommendation.
        \end{enumerate}
    \end{itemize}
    \item \textbf{Legal Case Analysis}
    \begin{itemize}
        \item \textbf{Scenario:} A lawyer requests recent legal precedents, statutory laws, and financial implications of a patent dispute.
        \item \textbf{Solution:}
        \begin{enumerate}
            \item The Orchestrator identifies subtasks: legal precedents, statutory research, and financial impact analysis.
            \item Agents query legal databases, legislative knowledge graphs, and financial systems.
            \item The results are synthesized into a comprehensive case summary.
        \end{enumerate}
    \end{itemize}
    \item \textbf{Financial Portfolio Management}
    \begin{itemize}
        \item \textbf{Scenario:} An investor asks for portfolio performance, market risks, and investment opportunities.
        \item \textbf{Solution:}
        \begin{enumerate}
            \item The Orchestrator splits the query into subtasks: performance metrics, risk analysis, and opportunities.
            \item Agents access portfolio databases, risk assessment tools, and market APIs.
            \item Results are combined into a personalized investment report.
        \end{enumerate}
    \end{itemize}
\newpage
    \item \textbf{Supply Chain Insights}
    \begin{itemize}
        \item \textbf{Scenario:} A logistics manager requests information on inventory levels, supplier performance, and shipment tracking.
        \item \textbf{Solution:}
        \begin{enumerate}
            \item The Orchestrator breaks the query into subtasks: inventory management, supplier analytics, and shipment tracking.
            \item Agents query inventory systems, supplier performance databases, and logistics APIs.
            \item Outputs are integrated into a detailed supply chain overview.
        \end{enumerate}
    \end{itemize}
\end{enumerate}

\subsubsection{Human-Augmented Agent}
A Human-Augmented Agent is an intelligent system designed to collaborate with humans by automating complex tasks while incorporating human oversight, feedback, or decision-making. These agents function as adaptive modules in larger systems, augmenting human capabilities by providing insights, generating recommendations, and performing tasks autonomously within predefined boundaries.

There are various patterns to implement Human-Augmented Agents based on application needs. For example:
\begin{itemize}
    \item \textbf{Human-in-the-Loop (HITL) Agent:} Integrates human feedback on decision status and environmental context to validate, refine, or override outputs generated by the agent.
    \item \textbf{Collaborative Agent:} Operates interactively with humans in real time, providing iterative suggestions or assisting in task execution.
    \item \textbf{Supervisory Agent:} Monitors processes, flags anomalies, and recommends corrective actions for human validation and intervention.
\end{itemize}

\begin{figure}[h!]
    \centering
    \includegraphics[width=\textwidth]{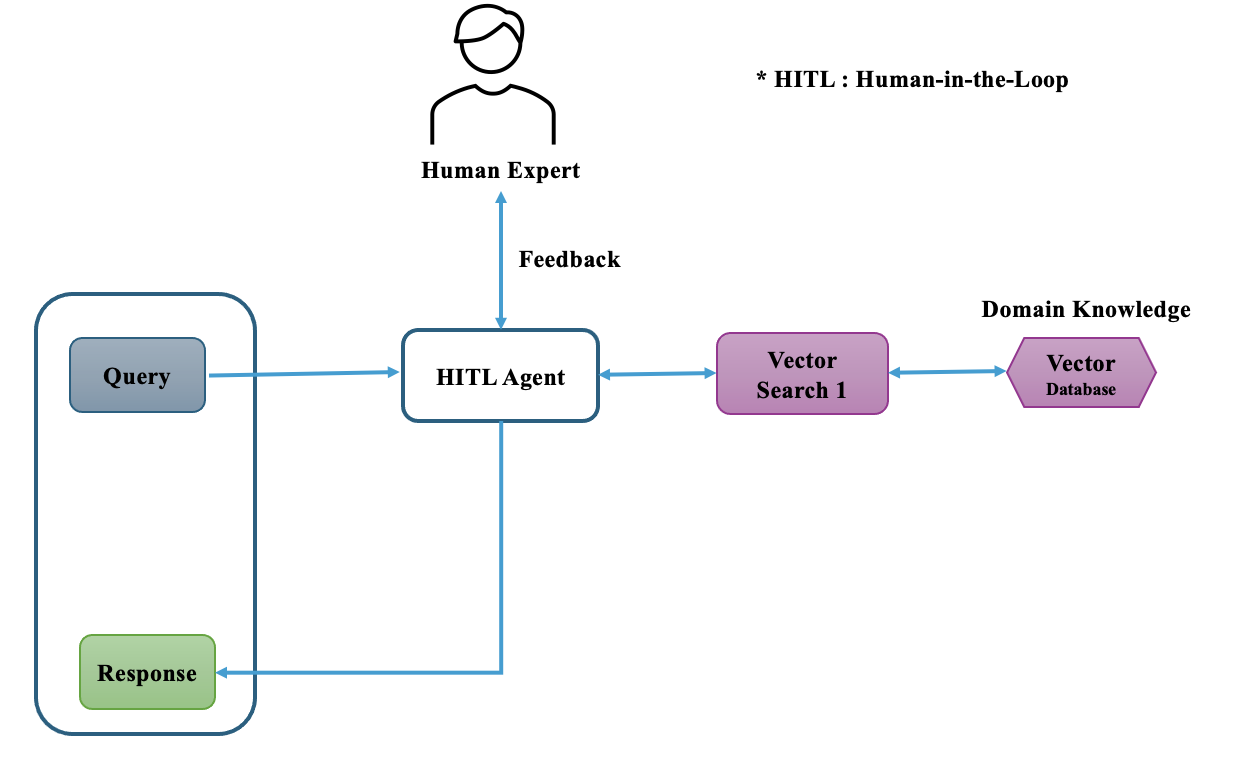}
    \caption{Human-in-the-Loop (HITL) Agent Pattern for Collaborative Decision-Making}
    \label{fig:hitl_agent}
\end{figure}

Figure \ref{fig:hitl_agent} illustrates the architecture of a Human-in-the-Loop (HITL) Agent Pattern, where the agent operates autonomously to process queries while integrating human expertise for validation and refinement.

\newpage
\paragraph{Key Components of the HITL Agent Workflow Pattern}
\begin{enumerate}
    \item \textbf{Query Input:} A user submits a query that is routed to the HITL Agent for processing.
    \item \textbf{Domain Knowledge Retrieval:} The HITL Agent uses a Vector Search mechanism to retrieve relevant information from a Vector Database, which contains domain-specific knowledge.
    \item \textbf{Response Generation:} Based on the retrieved information, the HITL Agent generates a preliminary response or decision.
    \item \textbf{Human Feedback:} The Human Expert reviews the agent's output, providing feedback on the decision status (e.g., approve, reject, modify) and offering additional contextual inputs if needed.
    \item \textbf{Feedback Loop:} The feedback from the human expert is integrated into the HITL Agent's reasoning process, allowing the agent to refine its understanding and improve future outputs.
    \item \textbf{Final Response:} Once validated or refined, the final response is delivered to the user.
\end{enumerate}

This pattern ensures that the agent operates autonomously within its domain but relies on human expertise to address high-stakes or context-sensitive decisions, enhancing reliability and adaptability. The feedback loop also enables the HITL Agent to learn and evolve based on human interactions, ensuring continuous improvement over time.

\paragraph{Practical Use Cases of Human-Augmented Agents:}
\begin{enumerate}
    \item \textbf{Healthcare Diagnostics and Treatment Planning}
    \begin{itemize}
        \item \textbf{Scenario:} A physician uses an AI system to assist in diagnosing rare diseases and formulating treatment plans.
        \item \textbf{Solution:}
        \begin{enumerate}
            \item The Human-Augmented Agent analyzes patient history, lab results, and clinical guidelines.
            \item It suggests potential diagnoses and treatment options, highlighting supporting evidence.
            \item The physician validates or refines the recommendations, ensuring the diagnosis aligns with patient-specific factors.
        \end{enumerate}
    \end{itemize}
    \item \textbf{Fraud Detection in Financial Systems}
    \begin{itemize}
        \item \textbf{Scenario:} A financial institution uses an AI system to monitor transactions for potential fraud or money laundering.
        \item \textbf{Solution:}
        \begin{enumerate}
            \item The agent flags suspicious activities based on predefined patterns and anomalies.
            \item A compliance officer reviews the flagged cases and validates whether they represent genuine threats.
            \item Feedback on false positives or new fraud techniques is shared with the agent to improve detection accuracy.
        \end{enumerate}
    \end{itemize}
    
    \item \textbf{Legal Document Review and Compliance}
    \begin{itemize}
        \item \textbf{Scenario:} A corporate legal team uses an AI system to ensure regulatory compliance in contracts and agreements.
        \item \textbf{Solution:}
        \begin{enumerate}
            \item The agent scans contracts to identify missing clauses, inconsistencies, or non-compliance risks.
            \item Lawyers validate and refine the flagged areas, tailoring them to specific regulatory requirements.
            \item The system learns from human feedback to improve future document reviews, ensuring faster and more accurate compliance checks.
        \end{enumerate}
    \end{itemize}
    \item \textbf{Real-Time Cybersecurity Monitoring}
    \begin{itemize}
        \item \textbf{Scenario:} Organizations use AI agents to monitor networks for cyberattacks or vulnerabilities.
        \item \textbf{Solution:}
        \begin{enumerate}
            \item The agent detects potential breaches or unusual activities (e.g., unauthorized access, malware).
            \item Security experts analyze flagged incidents to confirm the validity of the threat and determine mitigation actions.
            \item Feedback from resolved incidents helps the agent refine its threat detection and response capabilities over time.
        \end{enumerate}
    \end{itemize}
\end{enumerate}

\newpage
\section{Conclusion and Future Directions}

\subsection{Conclusion}
In this article, we explored the transformative power of agentic systems and their potential to address the dynamic and complex needs of modern industries. Beginning with the limitations of traditional SaaS platforms and the transition to context-aware systems, we established the foundational need for intelligent, adaptive solutions capable of operating in evolving environments. Vertical AI agents emerged as a critical innovation, offering operational advantages such as targeted domain expertise, real-time adaptability, and end-to-end workflow automation.

We examined the architecture and design of LLM agents, highlighting their core modules—Memory, Reasoning Engine, Cognitive Skills, and Tools, which equip them to process complex tasks in a scalable and domain-specific manner. The cognitive skills module was introduced as a key feature, enabling purpose-built models such as compliance monitors, responsible AI classifiers, and domain-specialized inference tools, ensuring agents operate responsibly and efficiently.

Expanding the scope to multi-agent and human-augmented systems, we showcased how these advanced frameworks integrate vertical intelligence to redefine software optimization, design, and automation. With their architectural flexibility and diverse applications, agentic systems have demonstrated their ability to revolutionize industries, driving operational efficiency and intelligent decision-making.

The transformative power of agentic systems and vertical intelligence signifies a paradigm shift in how businesses approach software and automation. By embedding contextual awareness and adaptability into intelligent agents, these systems enable unprecedented scalability, responsiveness, and ethical innovation. As industries continue to face complex challenges, agentic systems will play a pivotal role in shaping the future of intelligent workflows, offering groundbreaking opportunities for innovation and growth.

\newpage

\subsection{Future Directions}
Key future directions include:
\begin{itemize}
    \item Developing standardized frameworks to enhance interoperability and scalability.
    \item Expanding domain-specific intelligence for broader adaptability.
    \item Advancing human-agent collaboration to improve reliability and trust.
    \item Addressing ethical and regulatory concerns to ensure responsible use.
\end{itemize}

Agentic systems hold immense potential to revolutionize industries and tackle complex societal challenges. Addressing these priorities will unlock their full impact, driving innovation and delivering meaningful benefits across domains.

\bibliographystyle{plain} 
\bibliography{science_template} 

\end{document}